\documentclass[12pt,aps,preprint,a4paper,showpacs,nofootinbib]{revtex4}
\usepackage{amsmath}
\usepackage{graphicx}

\newcommand{\re}{\ref}
\newcommand{\be}{\begin{equation}}
\newcommand{\ee}{\end{equation}}

\newcommand{\la}{\label}
\newcommand{\ber}{\begin{eqnarray}}
\newcommand{\eer}{\end{eqnarray}}

\begin{document}
\title{Method to solve integral equations of the first kind with an approximate input}

\author{Victor D. Efros\footnote{v.efros@mererand.com}  
    }

\affiliation{European Centre for Theoretical Studies in Nuclear Physics and Related Areas 
(ECT*), Villa Tambosi, I-38123 Villazzano (Trento), Italy
\footnote{on leave from National Research Centre 
  "Kurchatov Institute",  123182 Moscow,  Russia}
}

\begin{abstract}
Techniques are proposed for solving integral equations of the first kind with an input known not precisely.
The requirement that the solution sought for includes a given number of maxima and minima is imposed.
It is shown that when the deviation of the approximate input from
the true one is sufficiently small and some additional conditions are fulfilled  the method leads 
to an approximate solution 
that is necessarily close to the true solution.
 No regularization is required in the present approach.
 Requirements on features of the solution 
at integration limits are also imposed. The problem is treated with the help of an ansatz proposed 
 for the 
derivative of the solution. The ansatz is the most general one compatible with the above mentioned requirements. 
The techniques are tested with exactly solvable examples.  
Inversions of the Lorentz, Stieltjes and Laplace integral 
transforms are performed, and very satisfactory results are obtained.  The method is useful, in particular, 
for the calculation of quantum--mechanical reaction amplitudes and 
inclusive spectra of perturbation--induced reactions in the framework of the integral 
transform approach. 
\end{abstract}

\bigskip


\maketitle

\section{Introduction}
   Inversion of integral transforms which values are known only approximately is a long--standing issue.
The situation is unfavorable when the solution that corresponds to an approximate input
differs from the true solution in a way that it
includes  extra narrow peaks or quick oscillations.
In these cases
small changes in the input 
correspond to changes in the solution that are not small.

But in such cases the numbers of maxima or minima of the approximate solution and of the true solution 
would differ from each other. Therefore, to cure the  situation it is suggested in the present paper
to seek for the approximate solution in the class of functions having the same number of maxima and minima as the true
solution has. In general this number is not known but often it  is not hard to guess it  as discussed below.
A general ansatz for the solution  compatible with a prescribed number of maxima and minima 
is proposed below for performing inversion. 

In Sec. 2 the proposed inversion method is described. 
It is applicable to a wide class of integral equations of the first kind.

In particular, the techniques are  aimed to increase ability of the integral transform approach 
used 
 to calculate amplitudes and 
inclusive spectra of perturbation--induced reactions in the
framework of nuclear physics. 
An outline of that approach is presented in Sec. 3.

In Sec. 4 the techniques are tested with exactly solvable examples. 
The Laplace, Stieltjes and  so called Lorentz \cite{ELO94}
integral transforms are inverted. 

In Sec. 5 the issue of convergence of the method to the true solution is investigated. The 
important point 
is that
the solution is sought for in a {\it restricted class} of functions. This ensures the convergence.
(A different case with such features exists, and in this case the solution belongs to a known compactum, 
see e.g. \cite{TA77}.)

\section{Inversion method}

The integral equation 
\be
\int_{E_{thr}}^\infty K(\sigma,E)f(E)dE=\Phi(\sigma),\qquad \sigma_1\le\sigma\le\sigma_2\la{inteq}
\ee
is studied. All the quantities are real, the solution $f$ is smooth and unique. 
The $K$ operator  is assumed to be continuous at the 
norm definitions specified below. 
The exact $\Phi$ is not known so that one is forced to deal with an approximation 
to $\Phi$ denoted below $\Phi_{appr}$. 
In the case of few--body calculations 
performed in the framework of the  approach outlined in Sec. 3 the accuracy of an input
$\Phi_{appr}$ is normally at a per cent 
level. (The notation for the lower integration limit in Eq.~(\re{inteq}) is used in connection with the problems addressed 
in Sec. 3 and 4.
In the same connection the upper integration limit is set to be infinite.) 

As it is known the solution $f$ sought for may be unstable with respect to small changes of the input.
In the literature, various regularization procedures  
were put forward to suppress the instability. (A rather complete review 
can be found in Refs. \cite{TA77,verlan}, see also e.g. \cite{gub,gr}.)   In particular, 
within the approach outlined in Sec. 3 the following regularization procedure 
was always used so far. 
The solution was approximated by the expression
\be 
f_N(E)=\sum_{n=1}^Nc_n\varphi_n(E,\alpha)\la{exp}
\ee
where $\{\varphi_n\}$ is a set of basis functions. The linear parameters $c_n$ and the non--linear parameter $\alpha$
were found via fitting  the quantity $Kf_N$ to $\Phi_{appr}$.   
The regularization was realized with the choice of the $N$ value. This value should be not too high to exclude 
unstable behavior.
At the same time it should be
not too low so that the approximation of the true $f(E)$ with the Eq.~(\re{exp}) type expression
would be sufficiently accurate. (See e.g. Ref. \cite{ELOB07}
for further details.) 
The problem is that the two conditions are compatible with each other only to a certain degree.
If the accuracy in an input $\Phi$ is not high enough reasonable
inversion may become impossible or tricky. The issue of the proper choice of the regularization parameter
does not arise in the method of the present paper since no regularization is required here.  

The techniques proposed below  
are designed to increase the accuracy of inversion at a given accuracy in  $\Phi$. It should be noted in this
connection that when  Eq.~(\re{exp}) is used another source of instability exists
besides  uncertainties in the 
input. This instability is related to arising
ill--conditioned systems of linear equations for  the expansion coefficients from Eq.~(\re{exp}).
The  influence of corresponding round--off errors  on the solution becomes considerable
 when $N$ increases even if the exact input is employed. In the framework of the method described below
the round--off errors are probably less important. 

It is shown in the last section that when small uncertainties in an input are of non--random nature and the 
 round--off errors in calculation are not substantial no 
other sizable deviations from the true solution besides 
narrow peaks of small strength  may
exist in the problem.  
Adopting that this is the case 
let us impose the requirement that the solution sought for includes a given number of maxima and minima.
Then one may conclude, see also the last section, that 
if the guess as to their number is correct 
then those narrow peaks are excluded and correspondingly the approximate solution is close to
the exact solution "almost everywhere".  

The only possible exception are the points of maxima and minima of the solution
where  extra narrow peaks might appear. 
If this improbable situation takes place these peaks are in general  to be simply removed. 
Indeed, true narrow peaks may appear only 
when there exist specific physical reasons for this,
such as resonant states. 
When the above requirement on the number of maxima and minima is imposed no regularization is required. 
Below the solution is sought for in the class of functions satisfying this requirement. 
 
In many cases incorrect guesses as to  the number of maxima and minima of the solution are  easily rejected.
Some maxima or minima may be very weakly pronounced. 
In such a case it may occur that approximate solutions with different numbers 
of maxima and minima   are very close to each other and lead to the fits of a similar quality.
Each of these approximate solutions is acceptable. Apart from this case,  
if the inversion is performed under the condition that the number of maxima and minima is less than the true one 
this necessarily should
lead to a fit of a lower quality. And if the inversion is performed under the condition that their number is 
higher than the true one this should lead either to a fit of lower quality or 
to unrealistically  narrow extra peaks in the solution with peaks positions and amplitudes
being unstable.

The inversion procedure is as follows. 
In the problems addressed 
in Sec. 3
the threshold behavior of $f(E)$ at $E\rightarrow E_{thr}$ 
is known.\footnote{This behavior is deduced from the known  behavior at  $E_\gamma\rightarrow E_{thr}$ of the matrix elements 
entering Eq.~(\re{f}). In this limit they include the $E_\gamma$ dependence as a factor.}
In particular, one has
$f(E_{thr})=0$. Also $f(\infty)=0$.
Let us rewrite Eq.~(\re{inteq}) in terms of $df/dE\equiv f'(E)$ integrating by part: 
\be
\int_{E_{thr}}^\infty {\tilde K}(\sigma,E)f'(E)dE=\Phi(\sigma).\la{inteq1}
\ee
(${\tilde K}(\sigma,E)=-\int K(\sigma,E)dE$.) 
Eq.~(\re{inteq1}) supplemented with the conditions \be f(E_{thr})=0,\qquad f(\infty)=0\la{bc}\ee 
is equivalent to Eq.~(\re{inteq}) at inputs $\Phi$ that lead to the values (\re{bc}). To satisfy the conditions (\re{bc})
let us impose the requirement
\be \int_{E_{thr}}^\infty f'(E)dE\equiv f(\infty)-f(E_{thr})=0\la{cond1}\ee
 and subsequently find 
the solution to Eq.~(\re{inteq})  e.g. as \mbox{$f(E)=\int\nolimits_{E_{thr}}^E f'(E')dE'$}. 
The solution to Eq.~(\re{inteq1})
with the condition (\re{cond1}) is unique. Indeed, otherwise there would exist more than one 
solution to Eq.~(\re{inteq})
constructed in the above way.

To be specific let us consider below the most frequent case when  at high $E$ 
values the solution $f(E)$ decreases  as power of $E$.  
(In the case of problems addressed in Sec. 3, one can see that in order to lead not to a power but to
an exponential decrease of $f(E)$ 
the inter--particle interaction in the coordinate or momentum representation should be
analytic in all the coordinates, or momenta, respectively. This is not the usual case.) Let $f(E)$ have
precisely $N$ maxima and minima. Let $\zeta_{thr}(E)$ be
 a monotonous
factor reproducing 
the threshold behavior of $f(E)$ at \mbox{$E\rightarrow E_{thr}$},
behaving as power of $E$ (e.g. as a constant) at \mbox{$E\rightarrow\infty$}, 
and arbitrary otherwise. 
The following ansatz for
$f'(E)$ is suggested,
\be
f'(E)=C\zeta_{thr}'(E)\left[\prod_{i=1}^N (E-E_i)\right]\frac{e^{\gamma(E)}}{[(E/{\bar E})+1]^{\beta}}.\la{anz1}
\ee
Here \mbox{$\zeta_{thr}'(E)\equiv d\zeta_{thr}/dE$}, 
$\bar E$ and $\beta$ are parameters, $\gamma(E)$ is a smooth function finite 
both at \mbox{$E=E_{thr}$} and  at \mbox{$E\rightarrow\infty$}, and e.g. $\gamma(E_{thr})=0$.  
The parameter $C$ determines the overall normalization. This  expression is the most general one
for the derivative of a function that has precisely $N$ maxima and minima, 
a given threshold behavior, and a power decrease at \mbox{$E\rightarrow\infty$}. (Inflection points do not require a special
consideration. A given power decrease is provided
by the choice of the parameter $\beta$.)  
The function
$\gamma(E)$ may be taken e.g. in the form 
\be
\gamma(E)=\frac{\Delta E/{\bar E}}{(\Delta E/{\bar E})+1}\sum_{k=0}^\infty \frac{c_k}{[(\Delta E/{\bar E})+1]^{k}}\la{alpha} 
\ee
with \mbox{$\Delta E=E-E_{thr}$}. Eq.~(\re{alpha}) allows the description of 
next--to--leading terms in $f'(E)$ when  both $E\rightarrow E_{thr}$ and $E\rightarrow\infty$.
The fitting parameters are then $C$, $\{E_i\}$, ${\bar E}$, $\beta$, and $\{c_k\}$. 

 To satisfy the requirement (\re{cond1})
it is convenient to express one of the $E_i$ values in (\re{anz1}) in terms of the other fitting parameters.  
In certain cases also the sum--rule condition 
\be -\int_{E_{thr}}^\infty E f'(E)dE\equiv\int_{E_{thr}}^\infty f(E)dE=S\la{cond2}\ee
 may   be imposed where the value of $S$ is known, see Sec. 3.
From this condition it is convenient to express $C$ in terms of the other fitting parameters. 

The remaining parameters are to be determined from the least--square fit 
procedure 
\be
||\Phi_{appr}-{\tilde K}f_M'||\equiv||\Phi_{appr}-Kf_M||={\rm min}.\la{cond}
\ee
Here $f_M'$ is an approximation   to $f'$ of the form (\ref{anz1}) 
 with $M$ parameters retained,
$f_M$ is the corresponding approximation to $f$, and the norm is defined as
\be ||F||=\left[\int_{\sigma_1}^{\sigma_2}w(\sigma)F^2(\sigma)d\sigma\right]^{1/2},\qquad w(\sigma)\ge0.
\la{defnorm}\ee

(When the equation of Eq.~(\re{inteq}) type is considered with $a$ and $b$ being the lower and the upper integration
limits and when $f(a)\ne0$ and/or $f(b)\ne0$ one may proceed in a similar way. One may rewrite the equation 
in the form of the equation for
the derivative $f'(E)$  and from that
 equation one obtains \mbox{$f'(E)=p(E)f(a)+q(E)f(b)+r(E)$} where
$p$, $q$, and $r$ are known functions. Then it is easy to get the relation \mbox{$f(E)=A(E)+\lambda B(E)$} where
$\lambda$ may be chosen to be $f(a)$, $f(b)$, or $f(c)$, $c$ being an arbitrary point, and 
$A$ and $B$ are the corresponding known functions. 
If there is no point $c$ at which $f(c)$ is known 
then the parameter $\lambda$ is to be determined from the initial integral equation.)

In simple cases  the main features of the solution $f(E)$  are determined by its threshold behavior, 
its decrease at large $E$, and by positions and amplitudes
of its maxima and minima. For example, let us consider here
the one--maximum or one--minimum case. The mentioned features of the solution may be reproduced in the
framework of the
ansatz (\ref{anz1}) without the factor $\exp[\gamma(E)]$.
Therefore 
one may think that
the solution is rather insensitive to the values of the parameters determining
$\gamma(E)$. 
In addition, in many cases the dependence  of 
the quantity \mbox{$||\Phi_{appr}-{\tilde K}f_M'||$} to be minimized on the parameters is rather smooth.
Then a good strategy to find the global
minimum of Eq. (\re{cond}) would be in the first stage 
to disregard the factor $\exp[\gamma(E)]$ in Eq. (\ref{anz1})
 and to seek for the minimum with respect to other parameters on a grid
that includes sufficiently large number of points.
Grids where one grid is put inside another one may be used.
The values of the parameters
thus obtained may be used as starting values at a subsequent conventional minimization.

To carry out the latter 
minimization a good choice are special codes for the least--square fit with non--linear
parameters    (e.g. Ref. \cite{rec}, Sec 15.5). Codes 
that use derivatives are preferable since the derivatives 
of e.g. the
expressions (\ref{anz1}) and (\ref{alpha}) with respect to fitting parameters   are readily calculated.

At calculating the norm in (\re{cond}) one replaces the integral with an integral sum. A good approximation to the solution
may be obtained even if the number of terms in this sum is insufficient to reproduce the integral.

Practical criteria of the quality of an approximate solution  $f_M$ thus obtained are both its stability 
with respect to increase of the number  $M$ of fitting parameters  at a given $\Phi_{appr}$ and 
its stability with respect to variations of $\Phi_{appr}$.
If a realistic
 estimate of $||\Phi-\Phi_{appr}||$ is available, say $||\Phi-\Phi_{appr}||>\epsilon_0$,  
then the criterion of stability with respect to increase of the number $M$ of fitting parameters may be replaced
with the condition that the quality of the fit, i.e. the value of \mbox{$||Kf_M-\Phi_{appr}||$}, 
is comparable with $\epsilon_0$.

After this work has been completed I got to know about the work \cite{mor}
where an approach having some common features with the present one was considered. 
The techniques of Ref. \cite{mor} are different from those of the present paper.  
The authors consider segments of monotonicity of the solution i.e. segments of a constant sign
of its derivative 
(and also of constant curvature of the solution). 
Discretizing the problem they reduce the corresponding minimum condition to the problem of the quadratic programming
with linear constraints. However, no systematic way was given in Ref. \cite{mor}
to find the borders of the monotonicity segments i.e. optimal positions of maxima and minima. 
Thus, unlike the present method, in Ref. \cite{mor} the whole problem  seems to
have not been solved. Besides, the issues related to 
the appearance of unrealistically narrow peaks discussed above (in relation to a guess as to the true number of maxima
and minima) and in the
last section (in relation to convergence of the method)
are not considered in  Ref. \cite{mor}.  

It may also be noted that 
the techniques  of Ref. \cite{mor} require  finding  the minimum with respect to many variables (like $f_i\equiv f(E_i)$)
with 
constraints. While in the present techniques finding the minimum with respect to only a few
fitting parameters without constraints is required.  
Furthermore, the behavior of the solution when it approaches 
the integration limits is not reproduced exactly 
in the techniques of Ref. \cite{mor}. Exact in--advance--reproduction   of this behavior
in the present techniques increases an overall accuracy
of the solution.

\section{Integral transform approach for calculating reactions in quantum mechanics}

The material of the present section
is used below to generate realistic inputs $\Phi_{appr}$.
Some features of the ansatz   of the preceding section for the solution $f(E)$ are also related to this 
material. 

The techniques of
the preceding section are designed, in partcular, for applications in the framework of the approach that is reviewed
below.
Only a brief outline of the approach is contained here. More  details
can be found in the reviews \cite{ELOB07,efr99}. The approach is advantageous, in particular, 
in problems with many open channels of various nature i.e. when energy is not low.

Its main features 
are the following. 
The dynamics calculations to be performed are 
bound--state type calculations. In the course of calculations 
there is no need  to consider reaction channels, as well as reaction thresholds. 
Reaction channels and thresholds come into play at merely the kinematics level only
 after a dynamics calculation is done.

Continuum spectrum states never enter the game. In place of them, "response--like" quantities 
of the type  
\be R(E)=\sum_n\langle Q'|\Psi_n\rangle\langle\Psi_n|Q\rangle\delta(E-E_n)+
\sum\!\!\!\!\!\!\!\int d\gamma\langle Q'|\Psi_\gamma\rangle\langle\Psi_\gamma|Q\rangle\delta(E-E_\gamma)\la{r}\ee
are basic ingredients of the approach. Here $\Psi_n$  are bound states and  $\Psi_\gamma$ are 
continuum--spectrum states. They  represent a complete set 
of eigenstates of the Hamiltonian $H$
of a problem. The subscript $\gamma$ denotes collectively a set of continuous and discrete 
variables labeling the states which is symbolized in the summation over integration notation.
The normalizations \mbox{$\langle\Psi_n|\Psi_{n'}\rangle=\delta_{n,n'}$} and
\mbox{$\langle\Psi_\gamma|\Psi_{\gamma'}\rangle=\delta(\gamma-\gamma')$} are assumed so that
\be \sum_n|\Psi_n\rangle\langle\Psi_n|+\sum\!\!\!\!\!\!\!\int d\gamma|\Psi_\gamma\rangle\langle\Psi_\gamma|=I,\la{i}\ee
$I$ being the identity operator.

In the method discussed the quantities $R(E)$ of Eq. ~(\ref{r})  are 
obtained not in terms of  the complicated states $\Psi_\gamma$ entering their definition but via a 
bound--state type calculation. And reaction observables are expressed in terms of $R(E)$ as quadratures.

Let us first explain the latter of these points. Consider strong--interaction induced reactions.
Denote \mbox{${\cal A}\phi_i(E)$} and \mbox{${\cal A}\phi_f(E)$} the antisymmetrized "channel free--motion states".
Here the subscript $i$ refers to the initial state of a reaction, the subscript $f$ refers to final
 states of a reaction, $\phi_{i,f}(E)$ are products of fragment bound states  and of factors describing their free
 motion \cite{GW}, and ${\cal A}$ denotes the operator realizing antisymmetrization with respect to 
 identical particles \cite{GW}. Denote \mbox{${\bar\phi}_i(E)={\cal A}(H-E)\phi_i(E)$} and
\mbox{$ {\bar\phi}_f(E)={\cal A}(H-E)\phi_f(E)$}.
One has $\bar\phi_{i}={\cal A}V_{i}^{res}\phi_{i}$ and $\bar\phi_{f}={\cal A}V_{f}^{res}\phi_{f}$, where 
$V_{i,f}^{res}$ are interactions between fragments in the initial and final states. Here it will be assumed that
these interactions are of a short range so that the long--range inter--fragment Coulomb interaction
is disregarded.\footnote{This restriction can be weakened or removed. This is done in part in \cite{S}.}  
The $T$ matrix determining the reaction rates  is  \cite{GW}
\be T_{fi}=T_{fi}^{Born}+\langle{\bar\phi}_f(E)|(H-E-i\epsilon)^{-1}|{\bar\phi}_i(E)\rangle,\la{t}\ee
\mbox{$\epsilon\rightarrow+0$}. Here $T_{fi}^{Born}$ is the simple Born contribution,
\[T_{fi}^{Born}=\langle\phi_f|{\bar\phi}_i\rangle=\langle{\bar\phi}_f|\phi_i\rangle,\]
and the main problem consists in calculating the second contribution in (\ref{t}) that includes
the Green function \mbox{$(H-E-i\epsilon)^{-1}$}. This contribution may be represented as
\be\int dE' R_E(E')(E'-E-i\epsilon)^{-1}\la{rint}\ee
where 
\be R_E(E')=\sum_n \langle {\bar\phi}_f(E)|\Psi_n\rangle\langle\Psi_n|{\bar\phi}_i(E)\rangle\delta(E'-E_n)
+\sum\!\!\!\!\!\!\!\int d\gamma
\langle {\bar\phi}_f(E)|\Psi_\gamma\rangle\langle\Psi_\gamma|{\bar\phi}_i(E)\rangle\delta(E'-E_\gamma).\la{resp}\ee
The quantity (\ref{resp}) is just of Eq.~(\ref{r}) structure  (with the \mbox{$E\rightarrow E'$} replacement). 
Thus, indeed, to calculate  
matrix elements of the $T$ matrix it is sufficient to have quantities of this structure.  Once they are available
 the integrations (\ref{rint}) are readily done. 

In order
to calculate a perturbation--induced reaction amplitude \mbox{$\langle\Psi^-_f|\hat{O}|\Psi_0\rangle$} where  
${\hat O}$ is a perturbation and $\Psi_0$ is an unperturbed initial state
the same is to be done 
with  the \mbox{${\bar\phi}_i\rightarrow {\hat O}\Psi_0$} replacement 
 in the above relations.

Now let us explain the above mentioned point on calculating the Eq.~(\ref{r}) type quantities.
It should  also be noted that such quantities may be of interest themselves representing observable response 
functions for inclusive perturbation--induced reactions.
Let us rewrite Eq.~(\ref{r}) as
\be R(E)=\sum_n R_n\delta(E-E_n)+f(E),\qquad R_n=\langle Q'|\Psi_n\rangle 
\langle\Psi_n|Q\rangle,\la{c2}\ee
\be f(E)=\sum\!\!\!\!\!\!\!\int d\gamma\langle Q'|\Psi_\gamma\rangle\langle\Psi_\gamma|Q\rangle\delta(E-E_\gamma)\la{f}.\ee
Calculation of the bound--state contributions $R_n$ can be  done separately, see also below. 
 The contribution (\ref{f}) 
includes an integral over few-- or many--body continuum states $\Psi_\gamma$ that are very complicated except for 
low energies, and the problem just lies in calculating this contribution.
If $E_{thr}$ denotes the threshold value for continuum state energies then $f(E)$ is different from zero at
\mbox{$E_{thr}\le E\le\infty$}.

An easy task is the sum--rule calculation. Using Eq.~(\ref{i}) one gets 
\be\int_{E_{thr}}^\infty f(E)dE=\langle Q'|Q\rangle- \sum_n R_n.\la{s}\ee
Obviously, the quantity (\re{s}) does not allow reconstruction of $f(E)$ itself. To achieve this goal, let us 
consider "generalized sums" of the form
\be \int_{E_{thr}}^\infty K(\sigma,E)f(E)dE.\la{ineq}\ee
Using Eq.~(\ref{i}) one obtains "continuous sum rules"
\begin{eqnarray}\int_{E_{thr}}^\infty K(\sigma,E)f(E)dE=
\sum\!\!\!\!\!\!\!\int d\gamma\langle Q'|\Psi_\gamma\rangle K(\sigma,E_\gamma)
\langle\Psi_\gamma|Q\rangle
\nonumber\\
=\langle Q'|K(\sigma,H)|Q\rangle-\sum_n R_nK(\sigma,E_n)\la{31}
\end{eqnarray}
where as above $H$ is the Hamiltonian of the problem and $R_n$ are defined in Eq.~(\re{c2}).
If one is able to calculate the quantity \mbox{$\langle Q'|K(\sigma,H)|Q\rangle$} entering Eq.~(\re{31}) then one
comes to the integral equation (\ref{inteq}) for $f(E)$ with
\be\Phi(\sigma)=\langle Q'|K(\sigma,H)|Q\rangle-\sum_nR_nK(\sigma,E_n).\la{ph}\ee
And at proper choices of the kernel $K$ one can
completely reconstruct $f(E)$ from this equation.

The presented approach to calculate reactions has been introduced in \cite{efr85}.\footnote{For observable
responses $R(E)$, i.e. in case of inclusive perturbation--induced reactions,
 a bound--state type
calculation 
of their integral transforms  has been
 suggested in \cite{efr80} in case of the Stieltjes transform
and in \cite{lapl} in case of the Laplace transform. Inversions of the transforms
were not considered in those works.}  The transforms with the kernels \mbox{$K(\sigma,E)=(E-\sigma)^{-p}$} where $p=1$ and 2
were employed. These are the Stieltjes transform and the generalized Stieltjes transform. 
Here $\sigma$ is chosen  taking real values lower than 
the continuum spectrum threshold and ranging outside
neighborhoods of the  discrete spectrum values $E_n$. In accordance with Eq.~(\re{ph}) the input in the integral equation is
\be\Phi(\sigma)=\langle Q'|\frac{1}{(H-\sigma)^p}|Q\rangle-
\sum_n \frac{R_n}{(E_n-\sigma)^p}.\la{stp} \ee
Denoting \mbox{$(H-\sigma)^{-1}Q={\tilde\Psi}$} and  \mbox{$(H-\sigma)^{-1}Q'={\tilde\Psi}'$} this can  also be written 
in the two respective cases as 
\be\Phi(\sigma)=\langle Q'|\tilde\Psi\rangle-\sum_n \frac{R_n}{E_n-\sigma}\quad {\rm where}\quad(H-\sigma)\tilde\Psi=Q,
\quad {\rm and}\la{p1}\ee
\be\Phi(\sigma)=\langle \tilde\Psi'|\tilde\Psi\rangle -\sum_n \frac{R_n}{(E_n-\sigma)^2}
\quad{\rm where}\quad (H-\sigma)\tilde\Psi'=Q',\quad (H-\sigma)\tilde\Psi=Q.\la{p2}\ee
The states $\tilde\Psi$ and $\tilde\Psi'$ are localized. Therefore the inputs $\Phi(\sigma)$ are indeed calculable with 
bound--state type methods. 

The transform with the "Lorentz kernel" \cite{ELO94} was intensively used. The procedure of Eq.~(\re{exp}) was used
for the inversion.
The kernel can be written as 
\be K(\sigma,E)=[(E-\sigma^*)(E-\sigma)]^{-1} \la{lkk}\ee   
where \mbox{$\sigma=\sigma_R+i\sigma_I$} is now complex.
The quantity $\Phi(\sigma)$ obtained in this case is  of Eq.~(\ref{p2}) form  with the replacement 
\mbox{$(E_n-\sigma)^{-2}\rightarrow
[(E_n-\sigma^*)(E_n-\sigma)]^{-1}$}.
One can also obtain the Lorentz input $\Phi(\sigma)$ from the dynamics equations that, like the Stieltjes case (\ref{p1}),
involve only the initial--state source term $Q$.  
For this purpose one rewrites the above mentioned $\Phi(\sigma)$ in the form \cite{efr99}
\begin{eqnarray}\Phi(\sigma)=(2\sigma_I)^{-1}\langle Q'|{\tilde\psi}_1-{\tilde\psi}_2\rangle-
\sum_n \frac{R_n}{(E_n-\sigma^*)(E_n-\sigma)}\nonumber\\
{\rm where}\quad 
{\tilde\psi}_1=(H-\sigma)^{-1}Q,\qquad
{\tilde\psi}_2=(H-\sigma^*)^{-1}Q.\la{lst}\end{eqnarray}
Both ${\tilde\psi}_1$ and ${\tilde\psi}_2$ are calculated from the initial state of a reaction.
Final states enter here via the known quantity $Q'$ i.e. as quadratures.
 
In the cases (\ref{p1}) and (\ref{p2}) it is convenient to calculate $R_n$ as the limits of the expression
\mbox{$-(E_n-\sigma)\langle Q'|\tilde\Psi(\sigma)\rangle$} and of the expression  
\mbox{$-(E_n-\sigma)^2\langle \tilde\Psi'(\sigma)|\tilde\Psi(\sigma)\rangle$} 
respectively at $\sigma$ tending to $E_n$. Here $\tilde\Psi$ and 
$\tilde\Psi'$ are the solutions to the corresponding inhomogeneous equations. In the Lorentz case
one can use a similar relation   with
both $\sigma$  and $\sigma^*$ tending to $E_n$ i.e. with \mbox{$\sigma_R\rightarrow E_n$}  and 
\mbox{$\sigma_I\rightarrow0$}.

Choosing the kernel $K$ as that of the Laplace transform one gets
\[\Phi(\sigma)=\langle Q'|e^{-\sigma H}|Q\rangle-\sum_nR_ne^{-E_n\sigma}.\]
This quantity is known to be calculable with the help of the Green function Monte--Carlo method at least
in the cases when at each  total angular momentum and parity values there exists not more than one bound state.\footnote{
At a sufficient accuracy in the calculated \mbox{$\langle Q'|e^{-\sigma H}|Q\rangle$} values 
the quantities $R_n$ and $E_n$ can  also be treated
as fitting parameters along with those entering $f(E)$.}

In the framework of no--core shell model calculations  other kernels $K$ may probably 
also be used in the present context,
see \cite{ELOB07,efr99}.

\section{Examples of inversion}

In this section Eq.~(\ref{inteq}) with \mbox{$E_{thr}=0$} and with
\be f(E)=\frac{4}{\pi E_0}\frac{\sqrt{E/E_0}}{[(E/E_0)+1]^4}\la{exact}\ee
taken as an exact  solution  is considered. 
Let us use \mbox{$E_0=20.7212603615$}, cf. below.  Approximate solutions 
 denoted $f_{appr}$ will be obtained below. 
The task here is to find out accuracy with which   the approximate solutions reproduce the exact one 
given by Eq.~(\ref{exact}). 
The minimum set of fitting parameters  to be used arises 
when  \mbox{$\exp[\gamma(E)]$} is disregarded in Eq.~(\ref{anz1}). In 
the present one--maximum case the corresponding ansatz is
\be f'(E)=\frac{C}{\sqrt{E}}\frac{E-E_1}{[(E/{\bar E})+1]^{\beta}}.\la{2par} \ee
The parameters $C$ and $E_1$ are expressed in terms of the two other parameters ${\bar E}$ and $\beta$ with the help of the 
relations (\ref{cond1}) and (\ref{cond2}). 
Then the
two parameters, ${\bar E}$ and $\beta$, are to be fitted. Below results pertaining to the approximation (\ref{2par}) 
are presented.

It is seen that the derivative  of the solution (\ref{exact}) lies in the class of 
the functions (\ref{2par}).
Therefore if the exact input is used then
the two--parameter ansatz (\ref{2par}) should lead to the exact solution up
to numerical inaccuracies. However, this is not the case when approximate inputs are employed.
Nevertheless, it occurs  
that the inclusion of additional parameters representing the quantity $\gamma(E)$ 
does not lead to  noticeable changes 
of the corresponding approximate solutions obtained.\footnote{In some cases changes in the approximate solutions 
due to inclusion of those parameters prove to be  even less than numerical uncertainties at 
finding the minimum. The program "frprmn" from \cite{rec}, Sec. 10.6, is used in the 
present calculations to search for the minima.}

Below some kernels $K$ of interest  are considered in Eq.~(\ref{inteq}). 
The inputs corresponding to those kernels are \mbox{$\Phi=Kf$}
where $f$ is the solution (\re{exact}).
Realistic approximate inputs $\Phi_{appr}$ related to those exact $\Phi$ are generated below and 
with these $\Phi_{appr}$ approximate solutions $f_{appr}$ are  subsequently obtained  using 
Eq.~(\ref{cond}). 

The average relative deviation of $f_{appr}$ from the exact $f$ is chosen as the criterion 
of the accuracy of a solution. I.e.
the quantity  
\be\chi_{solution}=\left[\frac{1}{N_1}\sum_{i=1}^{N_1}\left(\frac{f(E_i)-f_{appr}(E_i)}{f(E_i)}\right)^2
\right]^{1/2}\simeq\left\langle
|\Delta f/f|\right\rangle\la{cr}\ee
is adopted as such a criterion.  The sum goes
over a large number of $E_i$ values within an appropriate range of $E$. 
This range is taken to be $0\le E\le 42$ in all the cases, 
cf. Fig.~1  below.
 
The average relative deviation of $\Phi_{appr}$ from the exact $\Phi$  
\be \chi_{input}=\left[\frac{1}{N_2}\sum_{i=1}^{N_2}\left(\frac{\Phi(\sigma_i)-\Phi_{appr}(\sigma_i)}{\Phi_{appr}(\sigma_i)}
\right)^2\right]^{1/2}\simeq\left\langle
|\Delta \Phi/\Phi|\right\rangle\la{cf}\ee
is taken as the criterion of the accuracy of an approximate input. The sum goes
over a large number of $\sigma_i$ values within the ranges of $\sigma$ employed to solve the problem. 
These ranges are specified below. 
(The quantities $\Phi_{appr}(\sigma)$ have
no zeros in the cases considered.)

At performing the minimization  the weight function $w(\sigma)$ entering (\ref{defnorm}) is chosen to be 
\mbox{$1/[\Phi_{appr}(\sigma)]^2$}.
I.e. the average square of the  relative deviation of $Kf_{appr}$ from $\Phi_{appr}$ is minimized. 
The corresponding fit quality  is reported below as well. As the criterion of this quality the 
value
\be \chi_{fit}=\left[\frac{1}{N_2}\sum_{i=1}^{N_2}\left(\frac{\Phi_{appr}(\sigma_i)-(Kf_{appr})
(\sigma_i)}{\Phi_{appr}(\sigma_i)}
\right)^2\right]^{1/2}\simeq\left\langle
|(\Phi_{appr}-Kf_{appr})/\Phi_{appr}|\right\rangle\la{fit}\ee
is taken. Here $f_{appr}$ is the approximate solution obtained via the minimization procedure of Eq.~(\ref{cond}).

\subsection{Lorentz transform} 

The normalized Lorentz kernel (\re{lkk}) 
\be 
K(\sigma_R,E)=\frac{\sigma_I/\pi}{(\sigma_R-E)^2+\sigma_I^2}\la{lk}
\ee
is used. The parameter $\sigma_I$
 determines the width of the kernel. At \mbox{$\sigma_I\rightarrow0$} the kernel 
turns into \mbox{$\delta(\sigma_R-E)$}. 
The kernel
$\tilde{K}$ of the corresponding Eq.~(\ref{inteq1}) to be solved equals \mbox{$-\pi^{-1}\arctan[(E-\sigma_R)/\sigma_I]$}.

The corresponding input $\Phi(\sigma_R)$ is given in a range of $\sigma_R$ values. If the exact $\Phi(\sigma_R)$
is analytic in this range as in the applications of the preceding section then the solution corresponding to 
this $\Phi(\sigma_R)$ exists, it is unique, and 
it is independent of the chosen $\sigma_R$ range. 
Indeed,
$f(E)$ satisfies the equation that is obtained by the 
replacement of the lowest integration limit $E_{thr}$ with $-\infty$ if one sets
\mbox{$f(E)=0$} at \mbox{$E<E_{thr}$}. 
Since the kernel (\re{lk}) is analytic 
$f(E)$ thus defined satisfies  the latter equation also in the whole range  \mbox{$-\infty<\sigma_R<\infty$} 
with $\Phi(\sigma_R)$ continued analytically onto this range.  
And the equation thus obtained is a convolution equation having a unique solution.

Results of the present inversion method are compared below with those emerging from the Eq.~(\ref{exp})
procedure.\footnote{In the Lorentz case the latter
procedure was studied in \cite{ELO99,ALRS,BL,ELOB07}. In Ref. \cite{ALRS} some other approaches were also 
tested.} The basis sets with  the $E^{1/2}$
threshold behavior entering (\ref{exp}) used in almost all practical calculations performed so far
are the following\footnote{More involved basis functions were used in the recent paper \cite{Leid}.} 
\be \varphi_n(E,\alpha)=E^{n-1/2}e^{-\alpha E}\qquad{\rm and}\qquad \varphi_n(E,\alpha)=E^{1/2}e^{-\alpha E/n}.\la{bf}\ee
Results that were obtained with these two sets are of similar quality. Below the first of the sets is employed.
The corresponding calculations will be referred to as the  "standard inversion".

At searching the parameters 
the Lorentz transforms of the approximate solutions are calculated
numerically both in the standard inversion and in the present techniques cases.  
The same applies to inversion of the Stieltjes transforms below.

First let us present the inversion results for the case when  
the input $\Phi(\sigma)$ exactly
 corresponding to the solution (\ref{exact}) is employed.
While this input  can be calculated directly
it is convenient  to obtain it in another way aiming
to subsequently generate  \mbox{$\Phi_{appr}(\sigma)$}. Let us come to the solution (\ref{exact}) 
proceeding from  the representation of
Eq.~(\ref{f}) type. Let the states $\Psi_\gamma$  in (\ref{f}) 
 be the eigenstates 
of the one--particle free--motion Hamiltonian with the orbital momentum equal to zero.
In this case one may use $E_\gamma$ as the $\gamma$ variable in Eq.~(\ref{f})
and the summation sign is to be omitted.
Let us also set $Q=Q'$ in (\ref{f}).
The quantity $f(E)$ then equals $|\langle\Psi_E|Q\rangle|^2$. 
The corresponding wave functions   
$\Psi_{E}({\bf r})$, normalized to \mbox{$\delta(E-E')$} as in Eq.~(\re{i}), are 
\be \Psi_{E}({\bf r})=\frac{c}{\sqrt{4\pi}}\frac{\sin(kr)}{r},\qquad c=\left(\frac{2M}{\hbar^2k\pi}\right)^{1/2}.\la{psi}\ee
Here $M$ is the particle mass taken below to be the mass of the nucleon, 
and \mbox{$E=(\hbar k)^2/(2M)$}.
Let us also set 
\be Q({\bf r})=\frac{1}{\sqrt{4\pi}}e^{-\eta r}.\la{q}\ee 
Bound state contributions entering relations of Sec. 3 are to be omitted. 
This gives the expression (\ref{exact}) with
\mbox{$E_0=(\hbar\eta)^2/(2M)$} and with the normalization factor replaced by \mbox{$4/(\pi E_0\eta^3)$}. 
The sum rule value (\ref{cond2}) and (\re{s})
equals $1/(4\eta^3)$. Below it is set \mbox{$\eta=1$ fm$^{-1}$}.
If  energies are measured in MeV and distances 
in fm  then one gets the $E_0$ value listed above.

According to  the preceding section, see Eq.~(\ref{p2}) and below, 
the transform $\Phi$ corresponding to Eq.~(\ref{exact}) can be calculated as
\be \Phi(\sigma_R)=\int_0^\infty|\psi(r,s)|^2dr\la{ltt}\ee 
where $\psi$ is determined from the  equation
\be -\frac{\hbar^2}{2M}\psi''-s\psi=re^{-\eta r}\la{deq}\ee
with \mbox{$s=\sigma_R+i\sigma_I$}. The solution sought for is localized and satisfies the condition 
\mbox{$\psi(0,s)=0$}. In all the cases 
the inputs $\Phi$ or $\Phi_{appr}$ will be used in the range \mbox{$-2\le\sigma_R\le41.4$} with 
$\sigma_R$ measured in MeV units. (One may hope to obtain the solution $f(E)$ 
with a reasonable accuracy in the range \mbox{$0\le E\le E_{max}$} if the input of a similar
accuracy in the range \mbox{$-\sigma_I\le\sigma\le E_{max}+\sigma_I$} is employed, cf. also Fig.~1.)
 
The exact input deduced from Eqs. (\ref{ltt}) and (\ref{deq})
is calculated analytically. 
In Table~1 the results obtained with this input via the standard inversion are listed at the $\sigma_I$ value equal to 10~MeV.
At this $\sigma_I$ value the width of the kernel is not very different from that of the solution, cf. Fig.~1
below.  Standard inversions were usually done at such a condition. The results in Table~1 are presented for various
choices of the number $N$ of basis functions retained in the expansion (\ref{exp}). 

It is seen that, although the exact input is used,
the accuracy of the solution obtained is limited and the best accuracy attained is at the level of one per cent.
This happens despite the fact that the transforms of the basis functions (\ref{bf}) are calculated with high accuracy.
While the quality of the fit improves monotonically as $N$ increases this is not the case as to the solution 
obtained. When $N$ exceeds 9 the quality of the solution deteriorates and the instability due to tiny 
numerical inaccuracies
starts to develop.

\begin{table}
\begin{tabular}{|c|c|c|}
\hline
$N$&$\chi_{fit}$&$\chi_{solution}$\\
\hline
5&8.3$\cdot10^{-4}$&5.2$\cdot10^{-2}$\\ 
\hline
8&2.3$\cdot10^{-5}$&1.4$\cdot10^{-2}$\\ 
\hline
9&6.2$\cdot10^{-6}$&1.3$\cdot10^{-2}$\\
\hline
10&4.1$\cdot10^{-6}$&2.65$\cdot10^{-2}$\\
\hline
\end{tabular}
\caption{The inversion results. The exact input. The "standard inversion", \mbox{$\sigma_I=10$~MeV}.}
\label{tab1}
\end{table}

In Table~2 the results obtained with the method of the present work in the framework of the ansatz (\ref{2par}) are 
listed. Very different choices
of the width parameter $\sigma_I$ are considered. In all the cases the accuracy of the solution obtained is 
incomparable with that for the above
case of the standard inversion. 

\begin{table}
\begin{tabular}{|c|c|c|}
\hline
$\sigma_I$, MeV&$\chi_{fit}$&$\chi_{solution}$\\
\hline
2&1.4$\cdot10^{-7}$&1.6$\cdot10^{-8}$\\
\hline
10&1.45$\cdot10^{-7}$&2.1$\cdot10^{-8}$\\
\hline
100&7.5$\cdot10^{-8}$&3.3$\cdot10^{-7}$\\
\hline
\end{tabular}
\caption{The inversion results. The exact input. The new method, two fitting parameters.}
\label{tab2}
\end{table}

Now let us pass to the results at approximate inputs $\Phi_{appr}$. 
Let us proceed as follows. Eq.~(\ref{deq}) is a special case of the dynamic equation \mbox{$(H-\sigma)\tilde\Psi=Q$}
from (\ref{p1}) and (\ref{p2}). Usually at solving such equations expansions over basis functions are 
employed in few--body calculations 
 and
main inaccuracies in $\Phi_{appr}$
come from truncations of the expansions.  To model this situation let us  solve Eq.~(\ref{deq}) also in the form of 
a similar expansion.
The number of terms $N_0$ retained in the expansion will be chosen such that  $\Phi_{appr}$ will have the accuracy 
at a per cent level.
Specifically, let us use the approximation
\be \psi(r,s)\simeq\sum_{n=1}^{N_0}c_n(s)\varphi_n(r),\qquad 
\varphi_n(r)=[n(n+1)]^{-1/2}b^{-3/2}rL_{n-1}^2(r/b)\exp[-r/(2b)].\la{exp1}\ee 
Here \mbox{$b=0.3$ fm$^{-1}$} is the length parameter, and $L_m^2(x)$ are the Laguerre polynomials. 
The basis functions are orthonormalized.
The expansion coefficients $c_n$ are determined from the projection linear equations. The corresponding matrix elements 
entering the equations and the resulting
approximate inputs (\ref{ltt}) are calculated analytically. The same is done  below in the case of the Stieltjes
transform. 

First let us consider  again the \mbox{$\sigma_I=10$~MeV} case. Let us set \mbox{$N_0=10$} in Eq.~(\ref{exp1}). 
This leads to the corresponding $\Phi_{appr}$ that has a 
three per cent accuracy. The accuracy is defined as in  Eq.~(\ref{cf}), 
\be\chi_{input}=3.0\cdot10^{-2}.\la{3pe}\ee 

The results of the standard inversion are presented in Table~3. 
\begin{table}
\begin{tabular}{|c|c|c|}
\hline
$N$&$\chi_{fit}$&$\chi_{solution}$\\
\hline
2&8.0$\cdot10^{-2}$&0.22\\ 
\hline
3&2.6$\cdot10^{-2}$&0.12\\ 
\hline
4&2.6$\cdot10^{-2}$&0.13\\
\hline
5&1.35$\cdot10^{-2}$&0.71\\
\hline
\end{tabular}
\caption{The inversion results. An approximate input. The "standard inversion", \mbox{$\sigma_I=10$ MeV}.}
\label{tab3}
\end{table}
One sees that at a  three per cent accuracy 
of the input the best accuracy of the solution is twelve per cent and thus is unsatisfactory. (One should mention 
at the same time that
many good accuracy results  were previously obtained in such type problems  applying the standard inversion. 
Probably the accuracy of the input 
was
considerably higher than three per cent in those calculations.) Furthermore, it occurs that the approximate solutions at
 $N_{max}=3$ and $N_{max}=4$ are very close to each other. However, they are still not close to the true solution. Thus, 
stability with respect to the number of basis functions retained at performing the standard inversion is not sufficient 
for an approximate solution to be close to the true solution. Stability with respect to variations of $\Phi_{appr}$ is also
required.

Now let us pass to the result obtained  with the method of the present paper in the same case (see Eq.~(\ref{3pe}) 
and above it, 
\mbox{$\sigma_I=10$ MeV}).
One has 
\be \chi_{fit}=2.9\cdot10^{-2},\qquad\chi_{solution}=1.65\cdot10^{-2}.\la{si10}\ee
The accuracy of the solution obtained 
is even better than the accuracy (\ref{3pe}) of
the input in contrast to the standard inversion results. 

Perhaps the latter results might be somewhat improved due to use 
of basis functions decreasing as powers of $E$ at large $E$ values. This would require inclusion  more 
non--linear parameters in each of basis functions. Anyway,  the advantage of the present method 
is that in its frames there is no need to worry about the choice of the regularization parameter and 
instability does not arise 
 at all.  

Let us also list  the results obtained with the present method at other $\sigma_I$ values. Let us 
first consider the case
of a very narrow kernel, \mbox{$\sigma_I=2$ MeV}. In this case the decrease of $\psi(r)$ as $r$ increases 
is slow
with many oscillations present in its range of localization. Therefore many basis functions in (\ref{exp1}) are required
to reproduce $\psi(r)$  with a reasonable accuracy. With \mbox{$N_0=60$} in (\ref{exp1}) one gets 
\mbox{$\chi_{input}=2.6\cdot10^{-2}$}.
Using the corresponding $\Phi_{appr}$ one obtains with the present method
\be \chi_{fit}=2.6\cdot10^{-2},\qquad\chi_{solution}=8.6\cdot10^{-4}.\ee
The accuracy of the solution obtained is even considerably higher than that of the input in this case.

Let us now consider the opposite case of a very broad kernel, \mbox{$\sigma_I=100$ MeV}. Very few basis functions 
in (\ref{exp1}) 
are to be retained in this case to provide a reasonable accuracy of the input. At the number \mbox{$N_0=3$} 
of these basis functions one gets 
\be \chi_{input}=1.55\cdot10^{-3}.\la{s100}\ee
The standard inversion is 
absolutely inapplicable
in this case since the instability develops even for the lowest $N$ values in (\ref{exp}). 
However, the method of the present work allows to cope with this case as  well. 
Using the corresponding $\Phi_{appr}$ one obtains with this method
\be \chi_{fit}=4.8\cdot10^{-5},\qquad\chi_{solution}=7.7\cdot10^{-3}.\ee
The attained accuracy  of the solution  is still very reasonable.

The results for this case are displayed in Fig.~1. The dashed curve represents the exact solution and the dotted
curve represents the approximate one. They are practically indistinguishable. The solid curve represents the approximate input
employed at performing the inversion. It is practically  a constant. However, 
its tiny deviations from a constant prove to allow
reconstructing the solution with a very good accuracy. 
\begin{figure}[ht]
\centerline{\includegraphics*[scale=0.2]{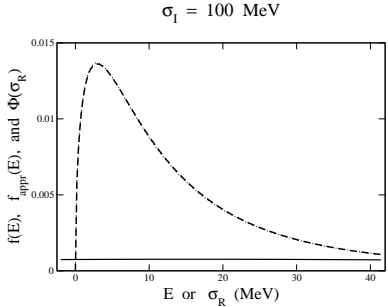}}
\caption{The inversion results. The new method, two fitting parameters. 
Accuracy of the approximate input is given by Eq.~(\ref{s100}). This input is represented by the solid curve.
The dashed curve and the dotted curve represent, respectively, the exact solution and the approximate one.}
\end{figure}

If at \mbox{$\sigma_I=100$ MeV}  the number $N_0$ of the basis functions in (\ref{exp1}) is increased up to e.g. 
\mbox{$N_0=10$} one gets 
\be \chi_{input}=9.3\cdot10^{-8}.\la{ss100}\ee
Using the corresponding $\Phi_{appr}$ one obtains in this last case with the present method
\be \chi_{fit}=2.1\cdot10^{-8},\qquad\chi_{solution}=2.1\cdot10^{-6}.\ee
This is to be compared with the values (\ref{si10}) obtained at the same $N_0$. 
One sees that, at the same number of basis functions 
employed to solve the dynamics equation, use of a very broad kernel is more preferable in the problem 
than use of a kernel with
the width comparable with that of the solution. This is due  to faster convergence with $N_0$ at solving the
equation (\re{deq}) of the  type of \mbox{$(H-\sigma)\tilde\Psi=Q$}, see the values in
Eqs. (\re{ss100}) and (\re{3pe}).

Comparison between results obtained from calculations at different $\sigma_I$ may serve for checking purposes.

\subsection{Stieltjes transform}

The Stieltjes kernel is
\be 
K(s,E)=(E-s)^{-1}.\la{sk}
\ee
Eq.~(\re{inteq}) is solved with $\Phi(s)$ given in a range of real negative $s$ values. 
If the exact $\Phi(s)$ is analytic in this range as in the applications of Sec. 3 then 
there exists a unique solution corresponding to this $\Phi(s)$  that is independent of this $s$ range. 
This follows from the fact that the Stieltjes transform is the iterated Laplace transform. 
Eq.~(\ref{inteq1})
with the kernel \mbox{$-\ln(E-s)$} is solved. 

The same procedure as above that involves Eqs. (\ref{psi}) and (\ref{q}) 
is used to calculate
the exact input and approximate ones. 
As it is seen from the preceding section, Eq.~(\ref{p1}), the exact input $\Phi$ that corresponds to the
solution (\ref{exact}) can be calculated in the Stieltjes case as
\be \Phi(s)=\int_0^\infty \psi(r,s)re^{-\eta r}dr\la{st}\ee
where $\psi(r,s)$ is the localized solution to Eq.~(\ref{deq}) at the condition \mbox{$\psi(0,s)=0$}. 
The same units and the same
$\eta$ value as above are used. The same procedure of expanding over the basis set (\ref{exp1}) is used to 
obtain approximate
inputs. 

At performing the inversion the inputs  in the ranges \mbox{$s_{max}\ge s\ge s_{min}$} are employed with various 
$s_{max}$ and with
\mbox{$s_{min}=s_{max}-41.4$ MeV}. To this aim, it is expedient that $|s|$ is confined to values not too much exceeding those
$E$ values at which $f(E)$ is substantially different from zero, cf. Fig.~1. At larger $|s|$ values the transform behaves 
as \mbox{${\rm const}\cdot s^{-1}$} and it
contains little additional information on $f(E)$. As to the choice of the $s_{max}$ value, the transform
exists for $s<0$ only and convergence of the approximate transform deteriorates when $s$ approaches zero.

At \mbox{$s_{max}=-2$ MeV}   in the case of the exact input one obtains
\be \chi_{fit}=3.45\cdot10^{-12},\qquad \chi_{solution}=3.9\cdot10^{-11}.\ee
In the case of an approximate input obtained with \mbox{$N_0=5$} in (\ref{exp1}) one has 
\mbox{$\chi_{input}=1.3\cdot10^{-2}$}.
Performing inversion with the corresponding $\Phi_{appr}$ one gets
\be \chi_{fit}=5.5\cdot10^{-3},\qquad \chi_{solution}=0.10.\ee
While this result is unsatisfactory, the increase of the number $N_0$ of the basis functions up to \mbox{$N_0=7$} leads to   
\be\chi_{input}=2.2\cdot10^{-3},\qquad
 \chi_{fit}=1.4\cdot10^{-3},\qquad \chi_{solution}=1.5\cdot10^{-2}.\ee
Let us mention that for obtaining the solution with a similar accuracy when the standard inversion is used the accuracy 
of the input
should be at the level of \mbox{$10^{-5}$--$10^{-6}$}.  

If at the same \mbox{$s_{max}=-2$}~MeV value one  increases further  the number of retained basis functions  up to 
\mbox{$N_0=10$}
one obtains  
\be \chi_{input}=1.2\cdot10^{-3},\qquad\chi_{fit}=1.4\cdot10^{-4},\qquad \chi_{solution}=1.1\cdot10^{-3}.\ee
If at the same \mbox{$N_0=10$} value one uses \mbox{$s_{max}=-10$}~MeV and \mbox{$s_{max}=-20$}~MeV  
one obtains, respectively,
\be s_{max}=-10\,\,\, {\rm MeV},\quad \chi_{input}=1.4\cdot10^{-7},\quad \chi_{fit}=9.9\cdot10^{-8},
\quad \chi_{solution}=2.2\cdot10^{-6}.\ee 
\be s_{max}=-20\,\,\, {\rm MeV},\quad 
\chi_{input}=2.3\cdot10^{-9},\quad \chi_{fit}=1.5\cdot10^{-9},\quad \chi_{solution}=1.0\cdot10^{-7}.\ee
Thus, despite the fact that the ratio \mbox{$\chi_{solution}/\chi_{input}$} becomes less favorable 
as $s_{max}$ decreases, the
accuracy of the solution itself increases. The reason for this is similar to that in the Lorentz transform case. 

Comparison of the results obtained from calculations at different $s_{max}$ values may serve for checking purposes.
Besides,  use of the Lorentz and Stieltjes transforms 
simultaneously for the same purposes is convenient.
 Indeed, the dynamic equations of Eq.~(\ref{deq}) type are the same in both cases. In addition, 
  in the Lorentz 
case it is in general convenient to calculate $\Phi$ from the Stieltjes 
 type relations, see Eq.~(\ref{lst}). 

\subsection{Laplace transform}

When
conventional regularization methods are used to invert the Laplace transform 
the instability problem is known in general to be severe. The Laplace transform  considered here is
\[\int_0^\infty e^{-zE} f(E)dE=\Phi(z),\qquad f(0)=f(\infty)=0.\]
Inversion is performed in the form
\[\int_0^\infty e^{-zE} f'(E)dE=z\Phi(z).\] 
The right--hand side values in the range  \mbox{$0\le z\le z_{max}$} with \mbox{$z_{max}\simeq 1.9304$ MeV$^{-1}$} are
used when performing the inversion. If ${\tilde E}$ is a typical scale pertaining  to the solution $f(E)$, cf. Fig.~1,
then at \mbox{$z\gg1/{\tilde E}$} the transform takes its asymptotic form 
and contains little additional information on $f(E)$. In the present case only minimization on grids is performed
without a subsequent conventional minimization. 

The transform $\Phi(z)$ of the exact $f(E)$ of Eq.~(\ref{exact}) is calculated numerically. 
The inversion with this accurate $\Phi(z)$ gives (using the two--parameter
ansatz (\ref{2par})) 
\[\chi_{fit}=2.6\cdot10^{-6},\qquad \chi_{solution}=8.0\cdot10^{-6}.\]

In the literature, in many--body calculations  inputs to  Laplace transforms
are obtained with the help of Monte--Carlo integrations. Therefore  
it will be assumed here
that inaccuracies of an input are of a random nature. Random inaccuracies may be modeled
in various ways. Below it is done as follows.  At each $z$ value the quantity $\upsilon\Phi(z)$ is added
to the accurate input $\Phi(z)$ where $\upsilon$ is a random variable
such that its mean value is zero.  Let us denote  $\tau^2$ its dispersion, i.e. $\langle\upsilon^2\rangle$. One may set 
$\upsilon=\tau\varrho$ where  $\varrho$ is a random variable with the unit dispersion. It is assumed that 
the distribution of $\varrho$ is the normal distribution and the values of  $\varrho$ are taken randomly 
in accordance with this distribution \cite{rec} at each $z=z_i$ value. 

The average relative error $\tau$ of the input
is taken to be $5.0\cdot10^{-2}$. Inversion with the approximate input $\Phi_{appr}(z)$ thus generated gives
\[\chi_{fit}=4.9\cdot10^{-2},\qquad \chi_{solution}=8.9\cdot10^{-3}.\] 
The value of $\chi_{fit}$ is such as one could expect. The result for $\chi_{solution}$ obtained
may be considered definitely good taking 
into account rather large relative error in the input.

\section{Convergence of the method}

In this section it is shown that  
if both the quantity \mbox{$||\Phi-\Phi_{appr}||$} is small and  saturation is achieved with respect to
the number $M$ of parameters  retained  e.g. in Eq.~(\re{alpha})  then the approximate solution $f_M$ 
from Eq.~(\re{cond}) is necessarily close  to the true $f$ everywhere except perhaps for
the points of maxima and minima of $f$. The case when the uncertainty \mbox{$\Phi-\Phi_{appr}$}
is not random is considered here so that one can speak of the norm \mbox{$||\Phi-\Phi_{appr}||$}. 
(In the above example the method works well also in the case of random uncertainty.)
Besides, the influence
of round--off errors on the minimization procedure (\re{cond}) is disregarded below. 

1. First, let us show the following. Let $f_M$ be the solution to Eq.~(\re{cond}). 
If  \mbox{$||\Phi-\Phi_{appr}||$} is small and   saturation with respect to
the number $M$ of parameters  determining $f_M$ is achieved then the quantity 
\mbox{$||\Phi-Kf_M||\equiv||K(f-f_M)||$}
is small as well. Namely,
\be
||K(f-f_M)||\le2||\Phi-\Phi_{appr}||+\epsilon_1\la{ap2}
\ee
where $\epsilon_1$ may be done arbitrarily small due to increase in the number $M$ of parameters retained.

Indeed,  one has
\be
||K(f-f_M)||\le||\Phi-\Phi_{appr}||+||\Phi_{appr}-Kf_M||.\la{1}
\ee
Let $f_M^0$ be an arbitrary  function of the structure of Eqs. (\re{anz1}) and (\re{alpha}) 
with the  same $M$ parameters retained as in the case of $f_M$.  
Due to Eq.~(\re{cond}) one can write  
\be||\Phi_{appr}-Kf_M||\le||\Phi_{appr}-Kf_M^0||\le||\Phi_{appr}-\Phi||+||\Phi-Kf_M^0||.\la{2}\ee 
Let us choose $f_M^0$ to
be arbitrarily close to exact $f$ taking $M$ value to be sufficiently large. 
 Since the $K$ operator  is continuous, 
at any \mbox{$\epsilon_1>0$} one can find  such $\delta$  in the corresponding relation \mbox{$||f-f_M^0||<\delta$} that 
one gets \mbox{$||\Phi-Kf_M^0||<\epsilon_1$}.  
Then combining Eqs. (\re{1}) and (\re{2}) one obtains Eq.~(\re{ap2}). (The above norm in the $f$ space
 may be defined differently from that in the $\Phi$ space.)

2. Now the following question is to be addressed. Suppose that \mbox{$||K(f-f_M)||$} is small. What are implications 
of this fact for the behavior of \mbox{$f-f_M$} itself? 
 
Below let us  define  the $K$ operator  as follows,
\be (K\psi)(x)=\int_a^bK(x,y)\psi(y)dy,\qquad c\le x\le d\la{abcd}\ee 
where \mbox{$K\psi\in L_2(c,d)$}, and the functions $\psi$ belong to the class of piecewise continuous functions.
The notation $||F||$ will refer to the $L_2$ norms below. Performing  a  change of variables
the norm of Eq.~(\re{defnorm}) may be
rewritten as the $L_2$ norm.
Let us also assume that $K$ is such that if \mbox{$||K\psi||=0$} then $||\psi||=0$
in the class of piecewise continuous $\psi$. 
(This condition is equivalent to the condition that the solution to the equation $Kf=\Phi$ is unique also in the
class of piecewise continuous  $f$.)

Consider the mapping \mbox{$Kz=u$}, where $z(y)$ are smooth uniformly bounded
functions,  \mbox{$||z||<Z$}, where 
$||z||$ is the \mbox{$L_2(a,b)$} norms of $z(y)$. 
Let  $||u||$ be the $L_2(c,d)$ norm of $u(x)$.

The following will be proved. 
At any $y_1$ and $y_2$ belonging to $[a,b]$ and at any $\epsilon>0$ one can find such $\delta$ that the relation
\be
\left|\int_{y_1}^{y_2}z(y)dy\right|<\epsilon\la{va}
\ee
holds true for all $z$ satisfying the condition 
$||u||<\delta$. (Here $u=u[z]$.) One can split the $[a,b]$ integration domain  from Eq.~(\re{abcd}) into segments
of a given length \mbox{$\Delta=y_2-y_1$}.
The above mentioned $\delta$ value may be found such that Eq.~(\re{va}) fulfills simultaneously
for all these  segments.

Eq. (\re{va}) provides a tool for establishing the local properties of the approximate solution.
Consequences for our purposes of this relation  will be discussed below after its derivation.
The above condition  that  \mbox{$z=f-f_{M}$} are uniformly bounded is equivalent to the condition that $f_{M}$ 
are uniformly bounded. 
If required, the latter condition may be
imposed at performing the minimization procedure.

To obtain the relation (\re{va})
let us write
\be ||u||^2=\int_a^b\int_a^bz(y){\cal K}(y,y')z(y')dydy'
\quad{\rm with}\quad {\cal K}(y,y')=
\int_c^dK(x,y)K(x,y')dx,\la{kt}\ee
\mbox{$a\le y,y'\le b$}.  In addition  it will be assumed here
that \mbox{$||{\cal K}||^2\equiv\int_a^b\int_a^b[{\cal K}(y,y')]^2dydy'$} is finite.  When $a$ or
$b$ is infinite this may   be not the case. Then for the purpose of the present reasoning
  the corresponding integration limit may be replaced with a large finite number. Indeed,
such a replacement does not influence physics, and the magnitude of the norm $||{\cal K}||$   will be of no relevance below.

For this purpose
in the above considered cases of the Lorentz, Stieltjes and Laplace transforms   one would need  to replace the infinite
upper integration limit with a finite $R$ value.\footnote{Below to employ expansions over continuum spectrum eigenfunctions
pertaining to these transforms could be an alternative.} It can be seen that in those cases
such a replacement leads merely to the replacement of the solution $f(y)$ with \mbox{$f(y)\theta(R-y)$}. It can be seen that 
the above mentioned
property \mbox{$||K\psi||=0\Rightarrow||\psi||=0$}, $\psi$ being piecewise continuous, is valid  in the case of those 
transforms both in their original
form and after the replacement of the upper integration limit with $R$.  

The kernel ${\cal K}$ is the Fredholm one and is symmetric. Therefore it possesses 
eigenfunctions and it is possible to number them \cite{tricomi}. Let us denote them $\phi_n$, $n=1,\ldots$, 
and let us denote
$\mu_n$ the eigenvalues,
${\cal K}\phi_n=\mu_n\phi_n$. Below let us choose $\phi_n$ to be orthonormalized.

Note that zero cannot be an eigenvalue of ${\cal K}$,  and $\mu_n>0$ at any $n$.
Indeed,  
according to Eq.~(\re{kt})  \mbox{$\mu_n=(\phi_n,{\cal K}\phi_n)=||K\phi_n||^2$}, and the kernel $K$ is such that 
\mbox{$||\phi_n||\ne0\Rightarrow||K\phi_n||\ne0$}.

The set of the eigenfunctions  of the kernel ${\cal K}$   is complete 
in the class of piecewise continuous functions in the sense of the approximation  in the  $L_2$ norm. 

Indeed, for any  function  $\psi$ orthogonal to all 
$\phi_n$  the equality ${\cal K}\psi=0$ is valid almost everywhere in $[a,b]$ \cite{tricomi}. Hence 
$(\psi,{\cal K}\psi)=0$ for such $\psi$ 
where $(\chi,\varphi)$ denotes the standard scalar product. But according to the definition (\re{kt}) the relation
$(\psi,{\cal K}\psi)=0$ yields \mbox{$||K\psi||=0$}  which was adopted above to be not possible  at
$||\psi||\ne0$ in case of the kernels $K$
under consideration. Thus if $\psi$ is orthogonal to all 
$\phi_n$ then $||\psi||=0$. I.e. the set $\{\phi_n\}$ is closed and therefore it is  complete. 
The set is thus infinite.

Let us pass now now to Eq.~(\re{va}).
 Let  $c_n$ be the Fourier coefficients
in the expansion of $z$ over $\phi_n$,
\be z(y)\sim\sum_{n=1}^\infty c_n\phi_n(y).\la{fou}\ee
  Let us represent the integral in Eq.~(\re{va}) as
\be \int_{y_1}^{y_2}z(y)dy=\sum_{n=1}^\infty c_nd_n\la{fou1}\ee
where $d_n$ are the Fourier coefficients in the expansion of \mbox{$\theta(y_2-y)-\theta(y_1-y)$} over $\phi_n(y)$. 
Eq.~(\re{fou1}) follows from the completeness property of the set $\{\phi_n\}$. Since all $||z||$ do not exceed 
some  $Z$ value
one has at any $n_0$ 
\be\left|\sum_{n=n_0+1}^\infty c_nd_n\right|\le Z\left[\sum_{n
=n_0+1}^\infty d^2_n\right]^{1/2}.\la{nu0}\ee
Taking this into account along with the fact that 
the function  \mbox{$\theta(y_2-y)-\theta(y_1-y)$} has a finite norm one sees that at any $\epsilon>0$ 
one can find such 
$n_0$ 
 that  the relation
\be \left|\sum_{n=n_0+1}^\infty c_nd_n\right|\le\frac{\epsilon}{2}\la{e1}\ee
is valid. 
The estimate is uniform with respect to  $z(y)$ out of the class considered. 

Furthermore, one has 
\be \left|\sum_{n=1}^{n_0} c_nd_n\right|\le S_1\le S_2\le S_3\la{ee}\ee
where
\[ S_1=\left[\sum_{n=1}^{n_0} c^2_n\right]^{1/2}(y_2-y_1)^{1/2},\]
\[S_2=\frac{1}{\mu_{min}(n_0)}\left[\sum_{n=1}^{n_0} \mu_n^2 c^2_n\right]^{1/2}(y_2-y_1)^{1/2},\] 
\[ S_3=\frac{||u||(y_2-y_1)^{1/2}}{\mu_{min}(n_0)},\]
and $\mu_{min}(n_0)={\rm min}(\mu_1,\mu_2,\ldots,\mu_{n_0})$. 
Let us set
\be||u||<\frac{\epsilon}{2}\frac{\mu_{min}(n_0)}{(y_2-y_1)^{1/2}}.\la{est1}\ee
Combining then Eqs. (\re{e1}) and (\re{ee}) one obtains that the absolute value of the 
quantity (\re{fou1}) does not exceed $\epsilon$, i.e. at arbitrary
$\epsilon$ one comes to Eq.~(\re{va}).

If the $[a,b]$ integration domain  in Eq.~(\re{abcd}) is split into the segments of the length of $\Delta=y_2-y_1$  then
at a given $\epsilon$ one can adopt the highest of the $n_0$ values pertaining to these segments
as the common $n_0$. The $||u||$ value in the estimate (\re{est1}) can be chosen the same for all 
the segments. Then Eq.~(\re{va}) will be valid for all the segments simultaneously.

3. The property (\re{va}) shows that at small $||Kz||$ the quantity $z$ may be not small  
only because of narrow peaks of high amplitude. And if the requirement is
imposed that the approximate solution $f_M$ has the same number of maxima and minima   
as the true $f$ has such narrow peaks of high amplitude are forbidden everywhere except 
for the points of maxima and minima of $f$. Therefore $z$ is necessarily small everywhere
except possibly for these points. 

Let us show this in more detail.  
Suppose that the $[a,b]$ integration domain is split into segments of the  length of $\Delta$. 
Below $\Delta$ is considered to be sufficiently small. Suppose that the  property (\re{va}) is fulfilled
for all the segments simultaneously. This imposes the following limitation on the behavior of $z(y)$. It varies within
the band $|z|\le\epsilon/\Delta$ and out of the band peaks with widths not exceeding $2\Delta$ are only 
permissible. (I.e. if a segment $[y_1,y_2]$  exists at which $z(y)>\epsilon/\Delta$
or $z(y)<-\epsilon/\Delta$ for all the $y$ values belonging the segment then necessarily $y_2-y_1\le2\Delta$.)
Indeed, otherwise the condition  (\re{va}) would be violated at some of the segments.


Let us exclude $\pm \Delta$ vicinities of maxima and minima of $f$ from the consideration (cf. the beginning
of Sec. 2 in this connection). Let us impose now
the requirement that the approximate solution $f_M$ has the same number of maxima and minima   
as the true $f$ has. This requirement puts limits on the amplitudes of the  above mentioned peaks possibly present 
beyond the band. 
These amplitudes cannot exceed the  \mbox{$|{\bar f'}|\cdot2\Delta+\epsilon/\Delta$} values where ${\bar f'}$ 
are average derivatives of the true solution in those segments (of the length not exceeding $2\Delta$) where the peaks are
located. Indeed, otherwise, due to these 
 peaks, the quantity \mbox{$f_M=f-z$} would include more maxima and minima than $f$ does. 
 But in this consideration $\Delta$ may be chosen  arbitrarily
small from the beginning. And when $\Delta$ is given $\epsilon(\Delta)$ may be taken arbitrarily small 
provided that $||K(f-f_M)||$ is sufficiently small. At this condition $|z|$ becomes smaller than any prescribed value.    

Thus in  the whole $[a,b]$ integration domain  with the points of maxima and minima of $f$ excluded the following
is valid.
If the mentioned requirement on the number of maxima and minima 
is imposed then there exist  $\delta_1$ values such that when \mbox{$||K(f-f_M)||<\delta_1$}, cf. Eq.~(\re{ap2}), 
the \mbox{max$|f-f_M|$} quantity   
 is necessarily
smaller than any prescribed value.

In conclusion, techniques to solve integral equations of the first kind with an approximate input
are proposed. 
It is proved that, at the  conditions and reservations listed above, the present method provides the solution
stable 
with respect  to perturbations of an input. No regularization is required at solving a problem in this way.
The fact that one need not deal with a regularization parameter  may in practice lead to 
a higher  accuracy. 
Some  inversion problems allowing comparison with an
exact solution of a rather simple structure are studied numerically. 
They include inversions of the Lorentz, Stieltjes and Laplace transforms
and involve systematic and random errors of an input. 
The
results prove to be  very satisfactory.

Acknowledgments of financial support are given  to
the RFBR, Grant No. 10-02-00718 and RMES, Grant No. NS-215.2012.2.

\end{document}